\documentstyle[preprint,aps]{revtex}
\draft
\begin{document}
\title{Local Spin-Gauge Symmetry of the Bose-Einstein 
Condensates in Atomic Gases}

\author{Tin-Lun Ho and V.B. Shenoy}
\address{Physics Department,  The Ohio State University, Columbus, Ohio
43210}
\maketitle

\begin{abstract}
The Bose-Einstein condensates of alkali atomic gases are spinor fields
with local ``spin-gauge" symmetry. This symmetry is manifested by a superfluid
velocity ${\bf u}_{s}$ (or gauge field) generated by the Berry phase of the
spin field.  In ``static" traps, ${\bf u}_{s}$ splits the degeneracy of the
harmonic energy levels, breaks the inversion symmetry of the vortex 
nucleation frequency ${\bf \Omega}_{c1}$, and can lead to {\em vortex ground
states}. The inversion symmetry  of ${\bf \Omega}_{c1}$, however, is not broken
in ``dynamic" traps.  Rotations of the atom cloud can be generated by 
adiabatic effects without physically rotating the entire trap. 
\end{abstract}

The recent discoveries of Bose-Einstein condensation in atomic gases of 
$^{87}$Rb\cite{Rb}, $^{7}$Li\cite{Li}, and $^{23}$Na\cite{Na}  have 
achieved a long sought goal in atomic physics. They have also provided
condensed matter physicists opportunities to study interacting Bose 
systems at a wide range of densities. 
The realizations of these condensates are made possible by the
invention of a number of special magnetic traps, which trap atoms with 
hyperfine spin $(F=2)$ maximally aligned with the local magnetic 
field ${\bf B}$. The reported Bose-Einstein 
condensations\cite{Rb}\cite{Li}\cite{Na}
 are found in these (adiabatic) spin states. 

An immediate question is whether these alkali condensates differ from
the familar $^{4}$He condensate in any fundamental way.
Unlike  the spinless $^{4}$He atoms, the trapped alkali atoms are in the 
$F=2$ hyperfine spin state. Their condensates are therefore spinors of the form 
\begin{equation}
<\hat{\psi}_{m}({\bf x},t)> = \zeta_{m}({\bf x},t) \Phi({\bf x},t) ,
\label{local} \end{equation}
where $\hat{\psi}_{m}$ is the field operator, 
$m$ is a label for $F_{z}$, $(-2\leq m\leq 2)$, $\Phi$ is a scalar, and 
$\zeta_{\mu}$ is a normalized spinor. Since 
the hyperfine spins are aligned with the magnetic field, $\zeta$ is 
given by $\hat{\bf B}\cdot {\bf F} \zeta = 2\zeta$, where 
${\bf F}$ is the hyperfine spin operator. The dynamics of $<\hat{\psi}_{m}>$ 
is therefore completely specified by that of the scalar field $\Phi$, 
as in $^{4}$He. One might then 
conclude that apart from extrinsic factors like density and external 
potential, there is no $intrinsic$ symmetry difference between $^{4}$He and 
alkali condensates. This is in fact the
starting point of all current theories, which model the alkali 
systems as interacting dilute $spinless$ Bose gases in harmonic 
potentials.  Within these models, the effective Hamiltonian for the scalar 
$\Phi$ has a global $U(1)$ gauge symmetry, as in $^{4}$He. 

The actual symmetry of the spinor field (eq.(\ref{local})), however, 
is much larger than $U(1)$. We call it $local$ spin-gauge 
symmetry. It represents that a gauge change $e^{i\chi({\bf x},t)}$ of 
$<\hat{\psi}_{m}>$ can be undone by a local spin rotation 
$e^{-i(\chi/F)\hat{\bf B}({\bf x},t)\cdot{\bf F}}$. 
Because of this symmetry, the exact form of the effective Hamiltonian of the 
scalar $\Phi$ is not that of $^{4}$He, but that of a neutral superfluid in a
velocity field ${\bf u}_{s}$, or an electron in a vector potential 
potential ${\bf A}$. The velocity (or gauge field) ${\bf u}_{s}$ arises from 
the Berry phase of the spin field $\zeta$. It is a direct
reflection of the underlying spin-gauge symmetry. 
The purpose of this paper is to discuss various forms of spin-gauge effects. 

As we shall see, the strength of the gauge field ${\bf u}_{s}$ 
is proportional to 
the gradient of the magnetic field ${\bf B}$. It is generally quite small
because ${\bf B}$ is generally fairly uniform in the trapping region. 
While these may be a justification of treating the alkali system like dilute 
$^{4}$He, we note that spin-gauge effects can be magnified rapidly by
variations of trap parameters and particle numbers.  As we shall see, 
despite its weakness, the effect of ${\bf u}_{s}$ can be observed over a wide 
range of trap parameters in both normal and superfluid phases. Moreover, 
in the condensed phase, particle interactions can reduce the excitation 
energy so much that the spin-gauge effect are significantly magnified.
In particular, when the excitation energy lies below that of spin-gauge
effect, the ground state will change abruptly to one reflecting this symmetry. 
Generally, spin-gauge effect increases as one moves away
from the center of the atom cloud. Understanding of this effect
is therefore important in view of the current effort to produce larger and 
larger clouds. It is also important to note that nonuniformity of the 
magnetic field, however small, is what leads to magnetic trapping. Spin-gauge 
effects are therefore intrinsic properties  of magnetically trapped
atomic gases. Our major findings are :

\noindent {\bf I}.  The adiabatic spin field $\zeta$ of the alkali system
generates a superfluid velocity ${\bf u}_{s}$. Its
vorticity ${\bf \Omega}_{s}$$=\frac{1}{2}{\bf \nabla}\times{\bf u}_{s}$
is specified by the magnetic field through a topological term, 
(eq.(\ref{curla}) below). 

\noindent {\bf II}. For cylindrical traps with static fields (or
simply ``static traps"), ${\bf u}_{s}$ generates a Coriolis force which 
splits the degenerate harmonic energy levels. This splitting is independent of 
particle interaction and can be observed 
in both normal and superfluid phases.

\noindent {\bf III}.  For static traps, the ``background" rotation $-{\bf
\Omega}_{s}$ breaks the inversion symmetry of the vortex nucleation 
frequency ${\bf \Omega}_{c1}$. 
For appropriate trap parameters, the system has 
a {\em vortex ground state} even in absence of external rotation. 

\noindent {\bf IV}. For cylindrical traps with time dependent fields, (or
simply ``dynamic traps"), ${\bf \Omega}_{s}$ has a quadrupolar structure. 
As a result, the inversion symmetry of ${\bf \Omega}_{c1}$ is restored. 

\noindent {\bf V}. Adiabatic effects furnish a simple means
to generate rotation of the atom cloud without physically rotating the entire
trap. 

To begin, we first discuss the effective Hamiltonian. For brevity, 
we shall call hyperfine spins ``spins". 
The Hamiltonians of the alkali systems are of the form $H=H_{s}+V$, where 
$H_{s}$$=\int d{\bf x}$$\hat{\psi}^{+}_{m}({\bf x})$
$[-\frac{\hbar^{2}}{2M}{\bf \nabla}^{2}$$-\mu_{a}{\bf B}({\bf x},t)\cdot
{\bf F}]_{mn}$$\hat{\psi}^{}_{n}({\bf x})$ is the single particle Hamiltonian, 
$M$ and $\mu_{a}=-\mu_{B}/2$ are the mass and magnetic moment of 
the atom, $\mu_{B}$ is the Bohr magneton, and the factor 1/2 is the
$g$-factor of the alkali atom.  
${\bf B}$ is a sum of magnetic field configurations which can be static 
or dynamic\cite{Rb}. $V$ is the two-particle interaction between the atoms.
To form a trap, the Zeeman energy $-\mu_{a}B$ (or their time average) 
must behave like a potential well.  If $\{ \zeta^{(n)}\}$ are the spin
eigenstates along $\hat{\bf B}$, ($\hat{\bf
B}\cdot {\bf F} \zeta^{(n)} = n \zeta^{(n)}$), ($-2\leq n\leq 2$), 
then the Zeeman energy
$-\mu_{a}{\bf B}\cdot{\bf F}$ reduces $U^{(n)}({\bf x}, t)
= \frac{1}{2}n\mu_{B}B({\bf x}, t)$ for  the states $\zeta^{(n)}$. 
If $\mu_{B}B$ is an attractive well, $U^{(n)}$ is confining (deconfining) 
for $n>0$ ($n\leq 0$). This means that spin-flips  
between $n>0$ and $n\leq 0$ states can cause atoms to leave the trap.
Since $V$ generally cause spin flips unless both atoms 
are in the  maximum spin state along the same quantization axis, 
(in which case spin flips are prohibited by angular momentum conservation), 
it depletes all but the ``adiabatic" spin states $\zeta^{(2)}$ 
in the trap. The resulting system is an interacting 
Bose gas with spins aligned with the local field  ${\bf B}({\bf x}, t)$.  

To construct a theory for the adiabatic spin states, 
we expand $\hat{\psi}_{m}$ in terms of the spin eigenstates
$\zeta^{(n)}$, 
$\hat{\psi}_{m}({\bf x},t)$$=\sum_{n=-2}^{2}$$\zeta^{(n)}_{m}({\bf x},t)$
$\hat{\phi}^{(n)}({\bf x},t)$. 
Expressing $\hat{\bf B}= \hat{\bf z}{\rm cos}\beta +
{\rm sin}\beta\left(\hat{\bf x}{\rm cos}\alpha + \hat{\bf y}{\rm sin}
\alpha\right)$, the explicit form of $\zeta^{(n)}$ is 
\begin{equation}
\zeta^{(n)}_{m} = <m|U|n>, \,\,\,\,\,\,\,\,\, 
U = e^{-i\alpha F_{z}} e^{-i\beta F_{y}}e^{-i\chi F_{z}}
\label{U} \end{equation}
where $F_{z}|n>=n|n>$. $\chi$ is arbitrary.  It is the gauge 
degree of freedom of the system, and is usually chosen to make the spinor 
$\zeta^{(n)}$ single valued.  The effective Hamiltonian ${\cal H}$  can be 
obtained by rewriting the equation of motion 
$i\hbar\partial_{t}\hat{\psi}_{m} = [\hat{\psi}_{m},H]$ in the form 
$i\hbar\partial_{t}\hat{\phi}^{(n)} = [\hat{\phi}^{(n)},{\cal H}]$. 
One then finds ${\cal H}={\cal H}_{ad}+{\cal H}_{nad}+
{\cal H}_{etc}$. ${\cal H}_{ad}$, referred to as the ``adiabatic" Hamiltonian, 
 contains $\hat{\phi}^{(2)}$ only. 
${\cal H}_{nad}$ is the spin-flip (or nonadiabatic) 
Hamiltonian which consists of cross terms between $\hat{\phi}^{(2)}$ and 
$\hat{\phi}^{(n\neq 2)}$. ${\cal H}_{etc}$ describes the transitions between
different $n\neq 2$ states and can be ignored. 
Denoting $\hat{\phi}^{(2)}$ and $\zeta^{(2)}$ as $\hat{\phi}$ and $\zeta$
respectively,  we have
\begin{equation}
{\cal H}_{ad} = \int d{\bf x} \hat{\phi}^{+}\left[
\frac{1}{2M}\left(\frac{\hbar{\bf \nabla}}{i} + M{\bf u}_{s}\right)^{2}
+ {\cal U} + {\cal W} \right]\hat{\phi} + {\cal V},
\label{Had}\end{equation}
where ${\cal U}$$=U^{(2)}$$=\mu_{B}B({\bf x})$, 
${\cal W}$$=(\hbar^{2}/2M)$$[\left|{\bf \nabla}\zeta\right|^{2}$
$+\left(\zeta^{+}{\bf \nabla}\zeta\right)^{2}]$
$-i\hbar\zeta^{+}\partial_{t}\zeta$, 
${\cal V}$ is the projection of $V$ onto the adiabatic spin states. It is of
the form ${\cal V}$$=\int V({\bf x}-{\bf y})$$\hat{\phi}^{+}({\bf x})$
$\hat{\phi}^{+}({\bf y})$$\hat{\phi}({\bf y})$$\hat{\phi}({\bf x})$, where 
$V({\bf x}-{\bf y})$ is a short range potential. 
The velocity ${\bf u}_{s}$ is  defined as 
\begin{equation}
M{\bf u}_{s}= \frac{\hbar}{i}\zeta^{+}{\bf \nabla}\zeta .
\label{us} \end{equation}
Eq.(\ref{Had}) describes a Bose fluid in a ``background" velocity
field $-{\bf u}_{s}$, or a charge $e$ system in a vector potential
${\bf A}$ if $M{\bf u}_{s}\equiv e{\bf A}/c$. Under a local spin 
rotation  exp$\left[ i\hat{\bf B}\cdot{\bf F}\chi({\bf x})\right]$, 
${\bf u}_{s} \rightarrow {\bf u}_{s} + (F\hbar/M){\bf \nabla}
\chi({\bf x})$, which is equivalent to a local gauge transformation 
$\hat{\phi}$$\rightarrow$ exp$(iF\chi({\bf x}))$$\hat{\phi}$. This 
is a reflection of the underlying spin-gauge symmetry of $\zeta$. 
The integral $\int_{C}{\bf u}_{s}\cdot {\rm d}{\bf s}$ is the Berry's phase 
of $\zeta$ around a loop $C$. It can be easily calculated from the vorticity 
(${\bf \Omega}_{s}$) of ${\bf u}_{s}$, which satisfies the Mermin-Ho
relation\cite{MH}, 
\begin{equation}
{\bf \Omega}_{s} = \frac{1}{2}{\bf \nabla}\times {\bf u}_{s} =
\left(\frac{\hbar}{2M}\right) \epsilon_{\alpha\beta\gamma}
\hat{B}_{\alpha}{\bf \nabla}\hat{B}_{\beta}\times {\bf \nabla}
\hat{B}_{\gamma}. 
\label{curla}
\end{equation}    
Eq.(\ref{curla}) shows that the spatial variations of ${\bf B}$ necessary
to produce the trapping potential will inevitably generate to a non-vanishing
superfluid velocity ${\bf u}_{s}$.

In the rest of this paper, we shall focus on the phenomena associated with
the adiabatic spin fields, (described by ${\cal H}_{ad}$ only). Nonadiabatic
effects will be discussed elsewhere\cite{HS}. Typically,  
nonadiabatic effects of the trap can be ignored if its ``Dirac center" is
sufficiently far away from the atom cloud. The ``Dirac center" is the point 
where ${\bf B}=0$ and that the unit vectors $\hat{\bf B}$ surrounding $D$ wraps
around the unit sphere $n$ times, ($n$ is a nonzero integer). If $D$ resides in
the cloud, the adiabatic spin field around $D$ will develop a line singularity
emerging from $D$, (a Dirac string), which will cause a lot of spin-flips. As
we shall see, increasing field gradients enhances spin-gauge effects but at the
same time moves $D$ closer to the cloud. The field parameters discussed below
are all within the range to keep the Dirac center sufficiently far
away from the cloud.

To be concrete, we consider ``static traps" of the form, 
(${\bf \nabla}\cdot{\bf B}={\bf \nabla}\times {\bf B}=0$), 
\begin{equation}
{\bf B}({\bf x}) = B_{o}\hat{\bf z} + G_{1}(x\hat{\bf x}- y\hat{\bf y})
+ \frac{G_{2}}{2}\left[\left(z^{2}-\frac{r^{2}}{2}\right)\hat{\bf z}-
z{\bf r}\right]
\label{Bex} \end{equation}
where ${\bf r}\equiv(x,y)$, $G_{1}$ and $G_{2}$ are the first and second 
order field gradients 
respectively.  Magnetic trap of the form eq.(\ref{Bex}) is similar to that used
in the $^{7}$Li experiment\cite{Li}. 
It is convenient to express the field gradients as $G_{1}\equiv 
B_{o}(\gamma/L)$, $G_{2}\equiv B_{o}/L^{2}$. The trapping potential 
${\cal U}$ in eq.(\ref{Had}) can then be 
expressed as 
\begin{equation}
{\cal U} = \hbar \Omega_{Zee} + \frac{1}{2}M\left[ \omega_{\perp}^{2}
r^{2} + \omega_{z}^{2}z^{2} \right]  + O|{\bf x}/L|^{4} . 
\label{Uex} \end{equation}
where $\hbar \Omega_{Zee} = \mu_{B}B_{o}$, 
 $\omega_{z}^{2} = \mu_{B}B_{o}/(ML^{2})$, 
$\omega_{z}/\omega_{\perp}=(\gamma^{2}-1/2)^{-1/2}\equiv \lambda$, 
$\gamma^{2}>1/2$. 
For later use, we denote the longitudinal and transverse width of the ground
state Gaussian of the harmonic well (eq.(\ref{Uex})) as $a_{z}$ and $a_{\perp}$,
where $a_{z}=(\hbar/M\omega_{z})^{1/2}$,  
$a_{\perp}=(\hbar/M\omega_{\perp})^{1/2}$. Typically, $a_{z}, a_{\perp}
<<L$.  It is straightforward to show that 
${\cal W}$$=(\hbar^{2}/2M)$$([{\rm sin}\beta{\bf
\nabla}\alpha]^{2}+[{\bf \nabla}\beta]^{2})$, where $\alpha, \beta$ are polar
angles of ${\bf B}$ as defined earlier. This term is smaller than the harmonic 
potential in ${\cal U}$ by a factor $(\gamma a_{\perp}/L)^{4}$ and can be
ignored in general. 

>From eq.(\ref{curla}), it is straightforward to show 
that [with $\chi=-\alpha$ in eq.(\ref{U})],
${\bf u}_{s}$$=\frac{2\hbar}{M}$$\left(1-B_{z}/B\right)$
${\bf \nabla}[{\rm tan}^{-1}(B_{y}/B_{x})]$, and 
\begin{equation}
{\bf u}_{s}=-\frac{\hbar}{M}
\left(\frac{\gamma}{L}\right)^{2}\hat{\bf z}\times {\bf r}
+O({\bf x}^{2}/L^{3}), \,\,\,\,\,\,
{\bf \Omega}_{s}= -\hat{\bf z} \frac{\hbar}{M}
\left(\frac{\gamma}{L}\right)^{2}
+O(|{\bf x}|/L^{3}).
\label{os} \end{equation}
Thus, for $|{\bf x}|< L$, spin-gauge effect generates a constant effective
``rotation" $-{\bf \Omega}_{s}$ along $\hat{\bf z}$.

An immediate consequence of ${\bf \Omega}_{s}$ is that it generates a Coriolis
force on the alkali system. This force can be detected 
by the applying an a.c. magnetic field along $\hat{\bf x}$, 
${\bf b} = b e^{-i\omega t}\hat{\bf x}$. This field will generate 
a term $(\mu_{B}b/2L) xe^{-i\omega t}$ in the effective Hamiltonian ${\cal
H}_{ad}$, as if a time dependent force ${\bf f}
= (\mu_{B}b/2L)e^{-i\omega t}\hat{\bf x}$ is present. 
It is easy to see that the equation of motion of the
center of mass in the $xy$-plane, ${\bf R}
= \int \hat{\phi}^{+}{\bf r}\hat{\phi}$, ${\bf r}=(x,y)$, assumes the form
\begin{equation}
M\frac{d^{2}{\bf R}}{dt^{2}} = - M\omega^{2}_{\perp}{\bf R}
+ 2M\frac{ d{\bf R}}{dt} \times
{\bf \Omega}_{s} + {\bf f} . \label{cmotion} \end{equation}
which has resonances at $\omega= \omega_{\perp}\pm \Omega_{s}$ (for
$\omega_{\perp}>>\Omega_{s}$). The degenerate clockwise and counterclockwise
harmonic modes $\omega_{\perp}$ are split by the Coriolis force. 
This splitting exists in both normal and superfluid phases, and can be easily 
shown to be independent of particle interactions. 

More pronounced effects can be found in the superfluid phase of alkali atoms
with positive scattering length $a>0$. 
Because of spin-gauge symmetry, the ground state energy functional becomes
\begin{equation}
{\cal E}(\Phi) = \frac{1}{2M}\left|\left(\frac{\hbar{\bf \nabla}}{i}
 +M{\bf u}_{s}\right)\Phi \right|^{2} +
({\cal U}+{\cal W}) \left| \Phi\right|^{2} +
\frac{2\pi \hbar^{2} a}{M}\left| \Phi\right|^{4}. 
\label{functional} \end{equation}
When ${\bf u}_{s}$  is small, eq.(\ref{functional}) can be written as 
${\cal E}(\Phi, {\bf u}_{s})$$= {\cal E}(\Phi, {\bf 0})$$-\Omega_{s}
{\cal L}_{z}$, (${\cal L}_{z}$$=-i\Phi^{\ast}\hat{\bf z}\cdot$
${\bf r}\times{\bf \nabla}\Phi$), which is the Hamiltonian density of 
a scalar superfluid in a container rotating with frequency 
$\Omega_{s}\hat{\bf z}$. Let $\Omega_{c1}^{o}$ denote the vortex nucleation 
frequency in the absence of spin-gauge effect (i.e. $\Omega_{s}=0$). 
Because of the ``background" rotation $\Omega_{s}$,
the actual vortex nucleation frequencies $\Omega_{c1}^{\pm}$ for 
vortices with $2\pi$ circulation around $\pm\hat{\bf z}$ will be 
$\Omega^{\pm}_{c1}= \Omega^{o}_{c1}\mp\Omega_{s}$. In particular, when
$\Omega_{s}\geq \Omega^{o}_{c1}$, hence $\Omega_{c1}^{+}\leq 0$, 
vortex ground state will emerge in the absence of external rotation. 

The value of $\Omega^{o}_{c1}$ has been studied for harmonic traps by a number
of authors\cite{Baym}\cite{String}. Using Thomas-Fermi approximation (TFA), 
which is good at large $N$ \cite{Baym}, Baym and Pethick have shown that
$\Omega^{o}_{c1}$ is reduced by particle interactions from its non-interacting
value $\omega_{\perp}$ as 
\begin{equation}
\Omega_{c1}^{o} /\omega_{\perp}= Q^{-2}{\rm ln}Q^{2}, \,\,\,\,\,\,
Q=R_{\perp}/a_{\perp}=(15\lambda Na/a_{\perp})^{1/5}. 
\label{oc1} \end{equation}
where $R_{\perp}$ is the transverse width of the condensate. Since
$\Omega^{o}_{c1}\propto N^{-2/5}$, $\Omega_{s}\propto N^{0}$, the inversion
asymmetry of the nucleation frequencies, 
$(\Omega^{+}_{c1}-\Omega^{-}_{c1})/(\Omega^{+}_{c1}+\Omega^{-}_{c1})$
$\approx$$\Omega_{s}/\Omega_{c1}^{o}$, increases as $N^{2/5}$. Thus, 
for sufficiently large $N$, the condition of vortex ground state 
$\Omega_{s}/\Omega_{c1}^{o}\geq 1$ can always be met. From 
eq.(\ref{os}) and eq.(\ref{oc1}), one finds that the ratio
$\Omega_{s}/\Omega_{c1}^{o}$ increases as the 
externally controllable parameters $N, G_{1}, B^{-1}_{o}$ increase. 

Figure 1 shows the ratio $\Omega_{s}/\Omega_{c1}^{o}$ as calculated from 
eq.(\ref{os}) and (\ref{oc1}) for $^{23}$Na, which has a positive scattering
length $a=4.9$nm. The asymmetry of the trap is set at $\lambda =1/2$. 
Four cases are considered : $N=5\times 10^{5}$ (broken line) and 
$N=5 \times 10^{6}$ (solid line); and $B_{o}=3$ and 5 Gauss, (denoted the
numbers 3 and 5 respectively). They are chosen to indicate the direction of
increasing spin-gauge effect as well as the conditions for vortex ground states. 
The field gradients considered extend to the
Tesla/cm range. Although 100 times higher than those in current 
experiments ($\approx 100$Gauss/cm), they are easily achievable using 
superconducting magnets. At present, the largest $N$ produced is $10^{5}$.
However, since $N$ has increased from $10^{3}$ to $10^{5}$ in last seven
months, it is conceivable that condensates with $N=10^{6}$ can be realized in
the near future.

>From figure 1, we see that for $N=5\times 10^{6}$, 
asymmetry ($\Omega_{s}/\Omega_{c1}^{o}$) up to 10$\%$ already occurs around
6500 Gauss/cm, and reaches 1 (i.e. vortex ground state) around 5 Tesla/cm. 
It should be noted, however, eq.(\ref{os}), (\ref{Uex}),
 (\ref{oc1}) neglected terms higher order in $|{\bf x}|/L$, 
which is justified only when $R_{z}\sim 2R_{\perp} <L$, where $R_{z}$ is the
longitudinal width of the condensate. 
For sufficiently large 
$G_{1}$ or $N$, this condition will fail, at which
point higher order terms in ${\cal U}$ and ${\bf u}_{s}$, as well as 
${\cal W}$ begin to contribute and that the system will lose 
cylindrical symmetry.
The values of $G_{1}$ at which $R_{z}=L$ are marked by circles on the curves 
in figure 1, indicating that eq.(\ref{oc1}) is only accurate 
to the left of the circle. 

To calculate $\Omega^{\pm}_{c1}$ in the regime $R_{z}\geq L$, we 
have calculated the energies ($E_{\pm}$ and $E_{0}$) 
of a $\pm 2\pi$ vortex
and the no vortex ground state using the full expressions
of ${\cal U}, {\cal W}$, and ${\bf u}_{s}$. The nucleation frequencies
$\Omega_{c1}^{\pm}$ are related to these energies as
$\Omega_{c1}^{\pm}\approx(E_{\pm}-E_{0})/\hbar$\cite{com1}. Our calculation are
performed within TFA, which in the present context amounts to replacing the 
kinetic energy by 
$(\hbar^{2}/2M)({\bf \nabla}\theta +M{\bf u}_{s}/M)^{2}|\Phi|^{2}$ and ignoring
the $(\hbar^{2}/2M){\bf \nabla}|\Phi|^{2}$ term. We have minimized this 
approximated 
energy subjected to the constraint of constant particle number 
$N=\int |\Phi|^{2}$, where $|\Phi|$ is the magitude of the order parameter of 
a $\pm 2\pi$ vortex (and the no vortex state) 
in the $E_{\pm}$ (and $E_{0}$) calculation.
Our results are shown in Figs.2a and 2b, which plot the ratio
$\eta^{\pm}=\Omega^{\pm}_{c1}/\omega_{\perp}$ as a function of $G_{1}$. 
We see that vortex ground states (i.e. $\eta^{+}=0$) emerge around 
5 Tesla for $N=5\times 10^{6}$. Even though 
the other cases considered have not reached vortex ground state, they show
strong broken inversion symmetry in the nucleation frequency, i.e. 
$(\eta^{+}-\eta^{-})/(\eta^{+}+\eta^{-})$ are close to or over 50$\%$. 
Finally, we note that the ratio $\Omega_{s}/\omega_{\perp}$, which 
describes the 
amount of energy level splitting in eq.(\ref{cmotion})  is of the
order of $10^{-3}$ for the range of parameters we considered, (which
is easily verified from the expression of $a_{\perp}$ and $\Omega_{s}$). 
This splitting, though small, is within the limit of detectability. 
Note also that the Dirac centers of the static trap
eq.(\ref{Bex}) are located at $(\pm L\sqrt{8\gamma^{2} +2},0, 2\gamma L)$,  
$(0, \pm L\sqrt{8\gamma^{2} +2},-2\gamma L)$. For the cases we considered, 
$\lambda=1/2$, hence $\gamma=\sqrt{4.5}$, these Dirac 
centers are quite far away from the
center of the cloud as $R_{z}\approx 2R_{\perp}$ is less than $2L$ for all 
cases considered. 

To further illustrate the spin-gauge effect, 
we consider dynamic traps like those in the Rb experiment\cite{Rb}, 
\begin{equation}
{\bf B}({\bf x},t) = B_{o}\left[ \hat{\bf n}(t) + {\bf b}({\bf x})\right], 
\,\,\,\,\,\,\, {\bf b}({\bf x})= 
\left({\bf r}-2z\hat{\bf z}\right)/L , 
\label{Bdyn} \end{equation}
where $B_{o}/L$ is the field gradient, and 
$\hat{\bf n}(t)= \hat{\bf p}{\rm cos}\omega_{o}t + 
\hat{\bf q}{\rm sin}\omega_{o}t$ is a unit vector rotating in a plane
perpendicular to $\hat{\bf l}$, and 
($\hat{\bf p}, \hat{\bf q}, \hat{\bf l}$) form an orthogonal triad. 
The adiabatic Hamiltonian ${\cal H}_{ad}$ is now periodic in time with 
frequency $\omega_{o}$. Expanding ${\cal H}_{ad}(t)$ in Fourier series 
of $e^{-in\omega_{o}t}$, the time averaged (i.e. $n=0$) term give rise to 
a static Hamiltonian of the form eq.(\ref{Had}) with ${\cal U}$ replaced by 
$\overline{\cal U} = \mu_{B}B( 1 + [{\bf b}^{2}+({\bf b}\cdot\hat{\bf
l})^{2}]/(4L^{2}) + ..)$. When $\hat{\bf l}=\hat{\bf z}$, 
$\overline{\cal U}$ has the 
cylindrical symmetric form eq.(\ref{Uex}) with
$\omega_{\perp}^{2}=\mu_{B}B_{o}/2ML^{2}$, $(\omega_{z}/\omega_{\perp})^{2}=8$. 
The rotational frequency $\omega_{o}$ has to fall in the range 
$\omega_{z}<\omega_{o}<\Omega_{Zee}$ for the spins to follow the local magnetic
field
adiabatically\cite{trap}. 
Using eq.(\ref{curla}), the
time averaged of the velocity field ${\bf u}_{s}$ and ${\bf \Omega}_{s}$ 
associated with eq.(\ref{Bdyn}) are found to be 
\begin{equation}
\overline{{\bf u}}_{s} = -\left(\frac{\hbar}{ML^{3}}\right)z\hat{\bf z}\times
{\bf r},
\,\,\,\,\,\,\,\,\, \overline{{\bf \Omega}}_{s} = 
\left(\frac{\hbar}{ML^{3}}\right)\left({\bf r}-2z\hat{\bf z}\right). 
\label{osbar} \end{equation}
Unlike the ${\bf \Omega}_{s}$ of the static trap, eq.(\ref{os}), 
$\overline{{\bf \Omega}}_{s}$ is a quadrupolar field which has mirror symmetry
about the xy-plane. As a result, $\Omega_{c1}^{+}=\Omega_{c1}^{-}$, 
which is similar to $^{4}$He but is entirely different from the static
trap\cite{bent}.
To verify this symmetry in the regime where $\overline{\Omega_{s}}$  
has similar strength as $\Omega_{s}$ in the static trap, the system has to be
rotated up to the critical frequency $\Omega_{c1}$, 
which can be as high as 100 rad/sec for $G=5000$ Gauss/cm, $B_{o}=5$ Gauss, and 
$N=10^{5}$. While it is impractical to rotate the entire trap 
at such high frequencies, we point out below a simple way to 
rotate the trapping potential using the adiabatic effects. 

Consider the case where $\hat{\bf l}$ deviates slightly from 
$\hat{\bf z}$, hence causing $\overline{\cal U}$ to deviate slightly from 
cylindrical symmetry. If $\hat{\bf l}$ precesses about $\hat{\bf z}$ 
with frequency $\omega_{p}$, $\omega_{p}<<\omega_{\perp}<\omega_{o}$, 
$\overline{\cal U}$ 
will rotate about $\hat{\bf z}$ with the same frequency. 
(The time average now is to be understood as averaging over times 
faster than $2\pi/\omega_{p}$). Since $\overline{\cal U}$ deviates only 
slightly from cylindrical symmetry, the corresponding vector fields 
$\overline{\bf u}_{s}$ and $\overline{\bf \Omega}_{s}$ are essentially given by
eq.({\ref{osbar}). Since the rotation ($\omega_{o}\hat{\bf l}$) and the 
precession ($\omega_{p}\hat{\bf z}$) of $\hat{\bf n}$ can be generated by 
electromagnetic means from sets of stationary coils, rotation of the
trapping potential can therefore be generated without physically 
rotating the entire trap. In a similar fashion, the potential ${\cal U}$ of the 
static trap can be made to rotate about $\hat{\bf z}$ by the application of a 
small magnetic field rotating in the $xy$ plane with frequency
$\omega_{p}<<\omega_{\perp}$\cite{rot}. Rotating the trapping potential in this
fashion allows one to study the inversion asymmetry of $\Omega^{\pm}_{c1}$ in
both low and high field gradient regime. 

We have thus established Statement {\bf I} to {\bf V}. 
Our discussions also show that spin-gauge effects assume different 
forms in different traps. It is therefore conceivable that they can be made 
more prominent at lower field gradients by other ingenious design of traps.
In a broader sense, the spin-gauge effect is only a subset of a much larger
class of phenomena associated with the topological excitations 
of the spin field, which are suppressed by the magnetic field in the current 
traps. 
Should it be possible to release part (or all) of the spins degrees of freedom 
in a new trapping design, the spin-gauge phenomena of the resulting condensate 
will be truly remarkable indeed.

TLH would like to thank Greg Lafyatis for discussions. This work is supported
in part by NSF Grant No. DMR-9406936.

\newpage

\noindent {\bf Caption}

\vspace{0.2in}

\noindent Figure 1. The ratio $\Omega_{s}/\Omega^{o}_{c1}$ as a function of 
field gradient $G_{1}$ for $^{23}$Na in a trap with asymmetry $\lambda=1/2$. 
The solid and broken lines represent
particle numbers $N=5\times 10^{6}$ and $N=5\times 10^{5}$ respectively. The
labels 3 and 5 on the figure denote $B_{o}=3$ and 5 Gauss respectively. 
On each curve, the region to the left (right) of the circle indicates the 
condition  $R_{z}<L (>L)$. 

\noindent Figure 2a (2b) shows the ratio
$\eta^{\pm}=\Omega^{\pm}_{c1}/\omega_{\perp}$
a function of field gradient $G_{1}$ for $B_{o}=3$ (5) Gauss. The meaning of 
solid and broken lines as well as the symbols ``3" and ``5" are identical to 
those in figure 1.

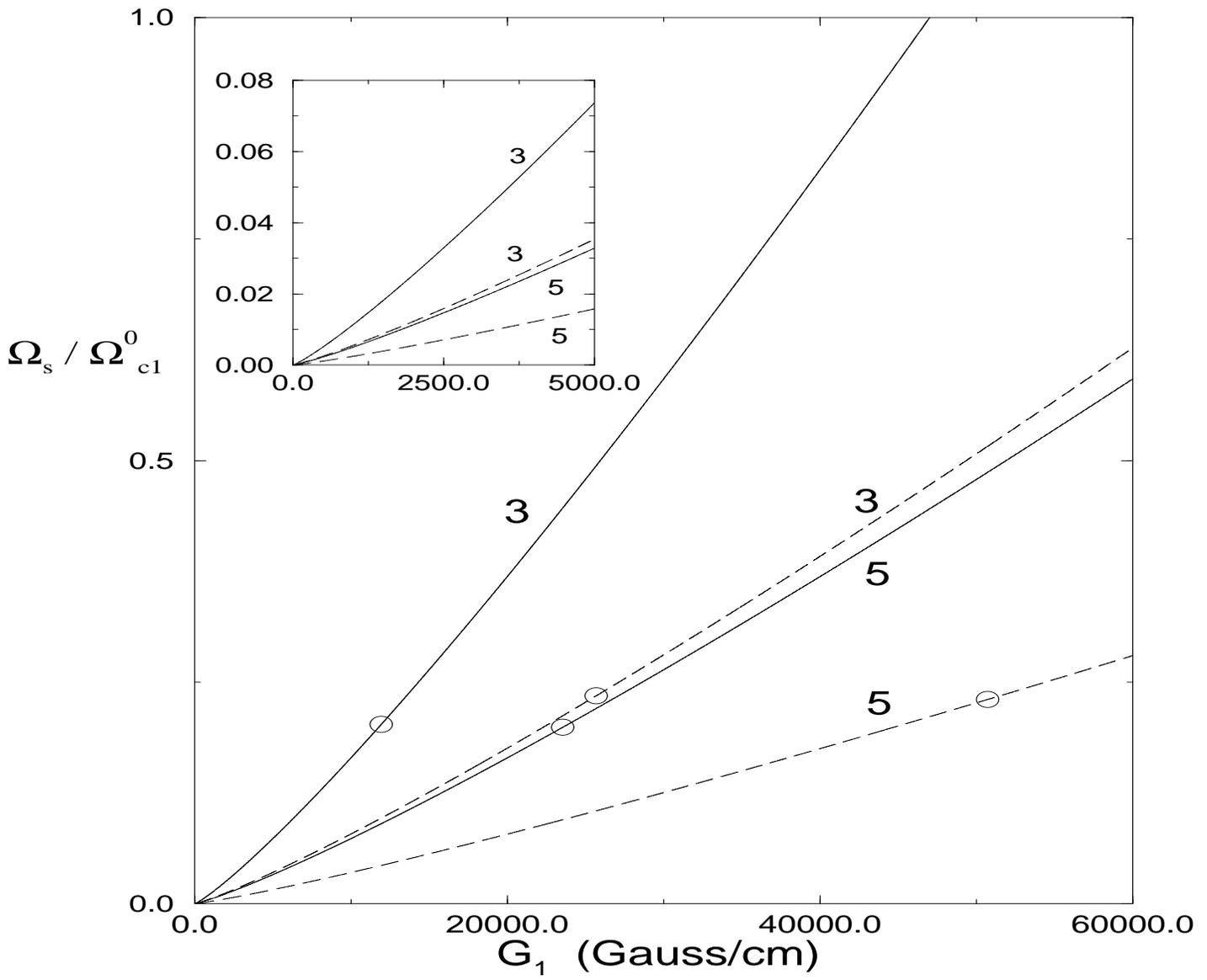
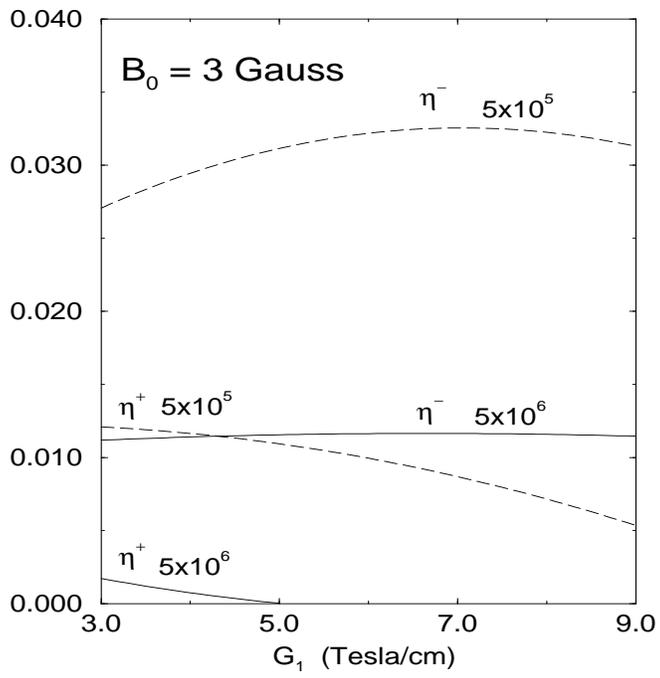
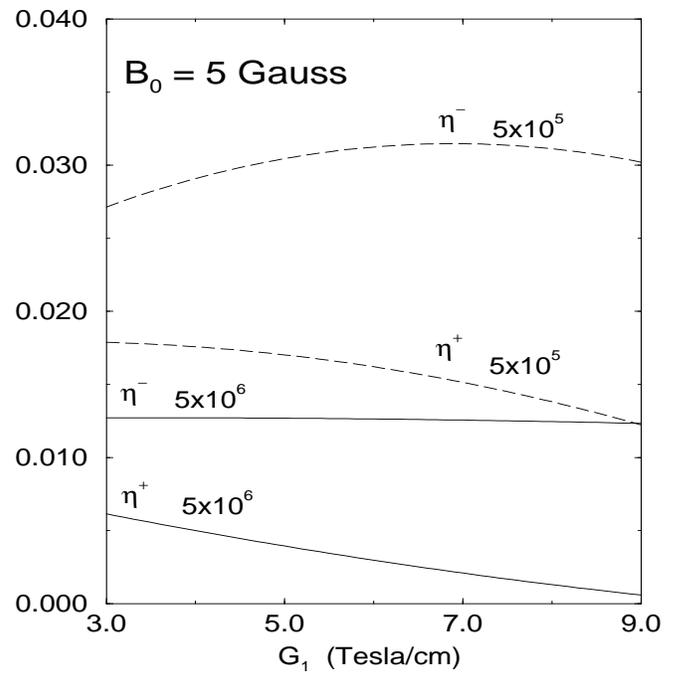
\end{document}